# Challenges and potentials for visible light communications: State of the art

Pranav Kumar Jha, Neha Mishra, and D. Sriram Kumar

# Challenges and Potentials for Visible Light Communications: State of the Art

Pranav Kumar Jha*, Neha Mishra and D. Sriram Kumar

*Department of ECE*

*National Institute of Technology, Tiruchirappalli, TN, India*

*jha.nitt.edu@gmail.com

**Abstract.** Visible Light Communication is the emerging field in the area of Indoor Optical Wireless Communication which uses white light LEDs for transmitting data and light simultaneously. LEDs can be modulated at very high speeds which increases its efficiency and enabling it for the dual purposes of data communication and illumination simultaneously. Radio Frequency have some limitations which is not at par with the current demand of bandwidth but using visible light, it is possible to achieve higher data rates per user. In this paper, we discuss some challenges, potentials and possible future applications for this new technology. Basically, visible light communication is for indoor application capable of multiuser access. We also design a very basic illumination pattern inside a room using uniform power distribution.

## INTRODUCTION

In the most recent couple of years, worldwide research in Visible Light Communication (VLC) utilizing LEDs for both light and information is at the pinnacle. This technology is useful mainly due to high transmission capacity and information rate, information security, no health hazards and low power utilization.

Nowadays, the incandescent bulb is gradually being eliminated as it has a low energy efficiency. LEDs are the best choice for illumination and also for communication purposes and it is fit for changing to various light power levels at a quick rate and the exchanging rate is sufficiently quick to be noticed by a human eye [3]. The information is encoded in radiating light in different ways and a photodetector at the receiver receives and decodes the modulated signal fulfilling the criteria of the dual purposes of illumination and communication. In the recent years, it has been demonstrated that VLC can achieve high information rates [1] as the communication encoded in visible light has uncommon significance relative to existing modes of wireless communication.

In the most recent two decades, confinements of Radio Frequency (RF) have been related to the exponential increment of traffic over mobile data networks. RF waves interfere with electronic devices and can infiltrate walls effectively prompts to the deterioration of signals and this attenuation in propagation limits information rates of proposed clients, however, visible light does not interfere with electronic devices. Since transmission does not bind to the expected region, the security of the connections might be traded off by listening stealthily malignant clients. VLC can give required coverage, visible light cannot infiltrate walls, so it has an inalienable connection security.

Visible light communication operating in the visible light range, which incorporates hundreds of terahertz of license free bandwidth shown in Fig. 1 [3], can correlate the RF based mobile communication systems in designing high capacity mobile data networks. Visible light cannot infiltrate through walls and objects which permits to generate small cells of LED transmitters with no inter-cell interference issues over the walls and parcels and feeds an inherent security for wireless data communications and can also increase the capacity of the available wireless channel.

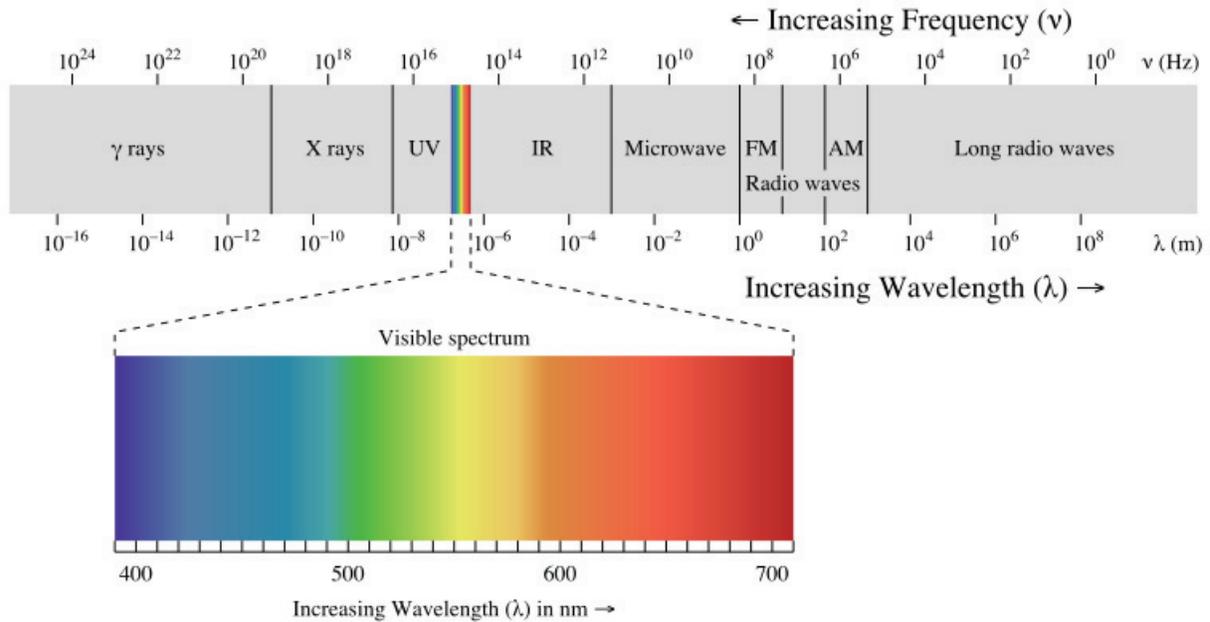

**FIGURE 1.** The human eye can sense the electromagnetic signals between the frequency range of 430 THz and 790 THz

## VLC OUTLINE

### Transmitter

LED is the most regularly used transmitters in visible light communication. It might be an LED or array of LEDs. The flow of current through the LEDs is controlled by a driver circuit and in turn brightness of transmitting light can be controlled. For communication purposes in both indoor and outdoor applications, white light is the most widely used forms of illumination because of color rendering [3]. Two types of LEDs are used to produce white light. By proper mixing of red, green and blue (RGB) light, white light can be generated. Another way to generate white light is to use a blue LED with yellow phosphor coating [3] where different intensities of white light can be generated by adjusting the thickness of the phosphor layer. RGB combination method of generating white light is not cost effective as it uses three separate LEDs costs higher than using blue LED with phosphor [3].

### Receiver

Either a photodetector or an imaging sensor (camera sensor) is used as receivers. The receiver consists of a photodetector and a setup for demodulation of the data. A camera sensor could also be used to receive the data sent from the LED array. An ideal photodetector should be sensitive to the wavelength interval analogous with the light, reliable and cost effective [3]. A photodetector should not be affected by the fluctuations in temperature. Any mobile device with a camera can be allowed to receive visible light for communication using image sensor. However, in its present form, it can just give exceptionally constrained throughput because of its low sampling rate. Then again, remain solitary photodetectors have appeared to accomplish fundamentally higher throughput [3].

## CHALLENGES

In general, optical communication is a system which is used every day by millions of users. Laser links are either space based or terrestrial point-to-point Free Space Optical Communications (FSO). Visible Light Communication is a completely new technology having the goal to utilize the lighting system for lighting and communication simultaneously. Though the growth of indoor wireless communication is growing very fast, Wi-Fi cannot keep up the increasing demands of the bandwidth. Wi-Fi can provide only limited data per user during heavy usage.

Later studies suggest that more than an eighty percent of wireless data communication engenders indoors. VLC system transmits information back and forth from infrastructure to mobile users for wireless data communication. To illuminate and to transmit energy in the form of data, LEDs are used frequently. It is not possible with the incandescent light sources or fluorescent light sources because it could not be modulated very effectively. LEDs are ubiquitous, non-coherent and very easy to modulate. LED can be modulated at a very reasonable rate. These devices are not designed for communications, only for lighting. It is not necessary to think about synchronizing carrier, phase or anything. The problem is their bandwidth is limited to a few megahertz (MHz) and they are not designed for high bandwidth.

## Improving Information Rate

For improving the rate of information, the most recent work is done in [4] where VLC guided-wave links based on the utilization of µLEDs and PAM schemes gives high-speed (>1 Gbps) optical interconnection in short-reach low-cost communication links [4]. Micro LEDs are much smaller in size, much faster rise time at a very low power level. Modulation, coding and pre-equalization [2] of transmitters are used for increasing the data rates. Suppose inside a hall, if everyone wants to connect to the internet to watch videos, handling of that might not be possible for Wi-Fi installed in that hall. With this technology, it is possible to use different levels of spatial multiplexing and spatial reuse of the resource in such a way that not everyone gets 1 Gbps, but it might be possible to supply everyone in that hall with enough bandwidth to watch their video and this is the problem addressed.

## MIMO

There is the possibility of significant multiplexing gain by having a multiple antenna supply signal to different receivers because receivers belong to own individual or multiple individuals [5]. Also, the possibility of using imaging receivers enables users to be in touch with the internet anywhere and anytime. For example, a camera is taking videos of lights and the lights could be encoded to downlink information directly through that and this possibility of using imaging receivers that had already built-in to all mobile devices already provides users to be connected to the internet endlessly.

## Modulation

The modulation techniques that are used in RF cannot be used for visible light communications. Flicker is a major problem as light is modulated at a very high speed and it may harm human eye. Dimming is also a problem for indoor environment where the amount of light needs to be controlled as per user's requirements. LED rise time and non-linearity are also a problem in modulation. Single carrier pulse based (MEPPM) and multicarrier (OFDM) modulation techniques are frequently used in visible light communications [6, 7]. LEDs are peak power limited and OFDM has high Peak to Average Power ratio. Conventional modulation for optical communication is pulse position modulation (PPM). Pulse amplitude modulation (PAM) cannot be used because of the flicker and dimming.

## Shadowing

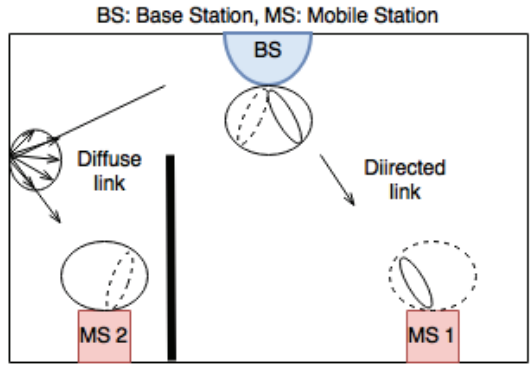

**FIGURE 2.** Directed and diffuse links

Where there is light, a single point is being illuminated by multiple light sources and then the shadowing problem translates to blocking one of many line-of-sight (LOS) links and for blocking all the LOSs, the concept of overprovisioning is used. If the adaptive modulation is used, it is possible to block some of those LOSs. Adaptive modulation and coding can be done to adjust to channel change. Fast accurate channel estimation is the key point for adjusting channel. When available, VLC can be used as an enhancement.

## Channel Modelling and Estimation

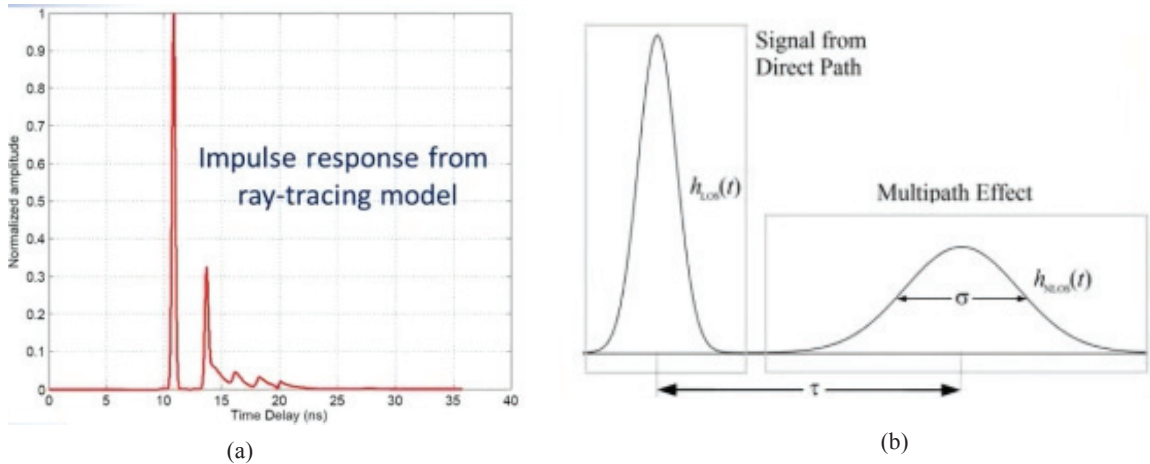

**FIGURE 3.** (a) Impulse response from ray-tracing model; (b) Multipath Effect

If a source and a receiver are used and the light goes LOS, it will reflect from all the walls and surfaces. In the Fig. 3(a), there are two LOSs and others is for reflections, a Ray-tracing model [9] is used to generate this impulse response. Here for the impulse response, a very short LOS Fig. 3(a) and a broad Non-LOS Fig. 3(b) is used and modulated as a Gaussian mixture. For address shadowing, rapid channel estimation can be used [10].

## Multi User Access

To provide data or information to the multiple users at a place, the supply of data should be in such a way that everyone gets connectivity for the resource available. Hence, for resource allocation so the resource or light source that are allocating power could be used efficiently, multiplexing could be done in different ways, and if it is orthogonal frequency-division multiplexing (OFDM), different frequency band for different users can be allocated. Frequency, time and code division multiplexing used can be same as RF but there is an addition that is spatial modulation [12].

## Uplink

To watch a video online selectively, requests should be generated and this requires uplink [8]. There are two ways in which it could be done. The first one is a stand-alone system and this prototype uses visible light, Infrared, RF or millimeter wave and if any of these is used for uplink, it can be completely stand-alone from the mobile back up into infrastructure. Visible light is not suitable otherwise the phone's LED and the phone will be switched on for requesting the video which is really not the purpose. Infrared is not used these days.

The second method is to converge system with others that are already deployed. Heterogeneous networks using Wi-Fi or cellular could be deployed as primarily downlink, but it freezes up a lot of Wi-Fi to service a lot of uplink and there is an asymmetry that the downlink tends to have a lot more demand than uplink. It might be possible to construct a balance between using more of the Wi-Fi for the uplink and less for the downlink

## APPLICATIONS

### Underwater Communications

VLC is a brilliant option for fast submerged communication contrasted with RF transmission, which is still to a great degree troublesome, and acoustic communication which comes about costly and have restricted the information rate.

### Hospitals

Wi-Fi is typically not allowed in hospitals because there are issues with health, interfering with electronic equipment. And if Wi-Fi is not there wireless is not there. It may be possible to provide some connectivity using VLC. This can be useful in critical situations which requires internet connectivity like monitoring patients from other locations without interfering with the other electronic equipment.

### Public Places

For outdoor application, LEDs are already deployed in the headlights and taillights of cars, streetlights and traffic signals. Using LED can provide potential to have communication going from traffic signal to the car, when there is an accident up ahead. This could save possible future accidents.

### Disaster Management

It may be possible to detect any minor cracks inside the aircraft engines and wings or any part of the aircraft, so that it could get fixed within time preventing air disaster. Fatigue failure is usually cannot seen by human eyes but using light it is possible to examine constantly, so that any possible cause for the accident could get fixed with less human effort. During heavy rain or fog, maneuvering the aircraft for landing is not an easy task, but it could be simpler with the help of this technology as light is already carrying information and information about the height and location of the aircraft can be exchanged automatically with the aircraft.

### Power Distribution inside a Typical Room Scenario

The indoor communication challenges, mostly depend on the power distribution of a typical room size. Fig. 4(a) shows a typical 4mx4mx3m room with a distance of 2.25m between the LEDs at the center of the top and the working space. The power distribution plots of the given scenario are determined for different irradiance angles for a 4x4 MIMO system and shown in Fig. 5 where different illumination patterns with wide and narrow coverage are obtained. The application of wide coverage is required for illumination purpose, whereas the narrow coverage targets point-to-point communication. In wide coverage, MIMO system suffers from co-channel interference rather than gaining mobility. In another case, the narrow coverage does not provide favorable regions for mobile users even though interference free. These factors, mainly affect the performance of MIMO systems. So, overcoming the challenge of tradeoff using unique power distribution pattern is a future direction [13].

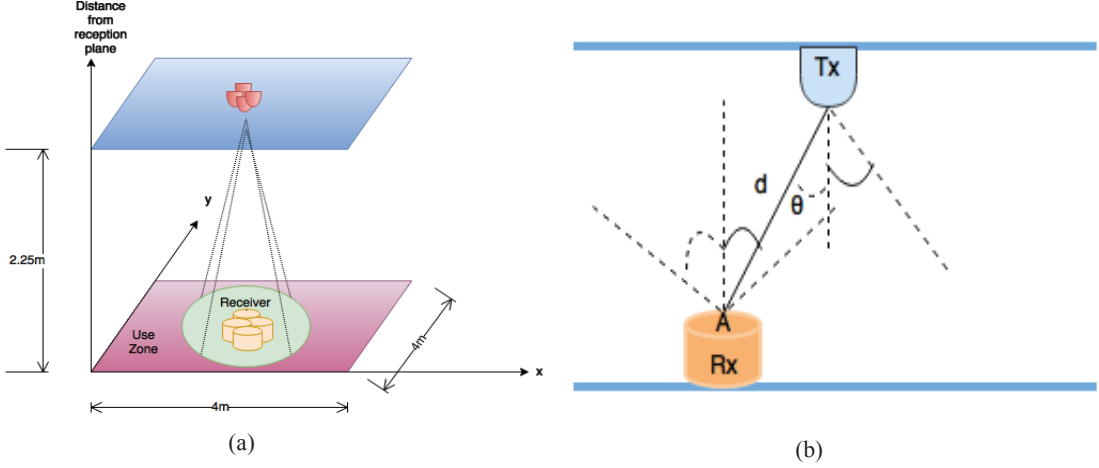

(a) (b)

**FIGURE 4.** (a) MIMO system in a typical room scenario; (b) System model with simulation parameters

The LED is accepted to have a Lambertian radiant intensity shown in the following equation [11, 13]:

$$R(\theta) = \frac{(m+1)\cos^m(\theta)}{2\pi} \quad (1)$$

Where the order of Lambertian emission $m$ sketched by the LED's semi-angle at half power $\theta_{1/2}$. The order of Lambertian emission described in the following equation [13]:

$$m = -\frac{\ln 2}{\ln(\cos(\theta_{1/2}))} \quad (2)$$

Where irradiance angle $\theta$ and incidence angle $\varphi$ respectively could be clearly shown in Fig. 4(b) [13]. Light passes from LED to receiver (Rx) through the LOS as the LEDs are situated at the center of the room. The received power can be given by the channel transfer function as [13];

$$h = \begin{cases} \dfrac{(m+1)A\cos^m\theta\cos\varphi}{2\pi d^2}, & \varphi \geq \varphi_{FOV} \\ 0, & \varphi < \varphi_{FOV} \end{cases} \quad (3)$$

Here the receiver (Rx) has an area $A$ and $d$ is the LOS distance between the LED and Rx. As shown in Eq. (3), the incidence angle $\varphi$ ought to be not as much as the field-of-view (FOV) incidence angle $\varphi_{FOV}$, else, the channel transfer function is zero [13].

**Table**

**TABLE 1.** Simulation parameters

| Space between LEDs (In meters) | Irradiance angle (In degrees) | Space between detectors (In meters) |
|---|---|---|
| 1.5 | 4 | 0.5 |
| 1.5 | 5 | 0.5 |
| 1.5 | 7 | 0.5 |
| 1.5 | 8 | 0.5 |

MATLAB 2015b is used for the simulation purpose and the received power distribution plots for different irradiance angles are presented in the figures below: **FIGURE 5.**

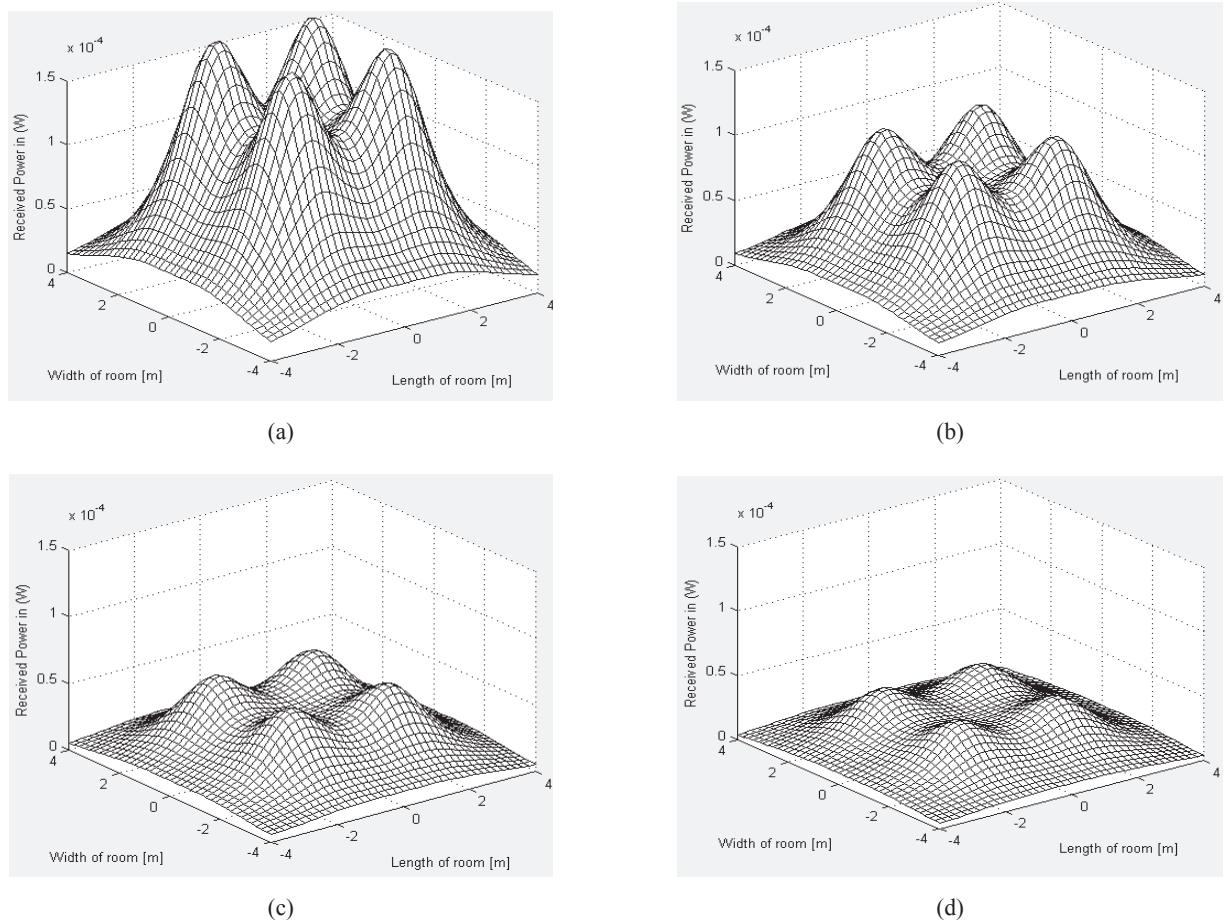

**FIGURE 5 (a)(b)(c)(d)**

## CONCLUSION

Different challenges and potentials for visible light communication along with the possible future applications has been discussed. A brief power distribution simulation has been done to show the different illumination patterns which give tradeoff between illumination and communication. It has been concluded that wide coverage suits for illumination where communication happens with high interference. To overcome the issue, a narrow coverage illumination pattern has been designed for 4 degree LED semi-angle, which suits best for fixed point-to-point line-of-sight indoor communications.